\documentclass[aps,prb,groupeaddress,supperscriptaddress,showpacs,footinbib,twocolumn,showkeys,final]{revtex4}
\usepackage{graphics,graphicx,color}
\def\tauphi{$\tau_{\phi}$ }
\def\lphi{$L_{\phi}$ }

\def\lso{$L_{so}$ }

\begin{document}

\title{Electron coherence at low temperatures: The role of magnetic impurities}
\author{Laurent Saminadayar$^{1}$, Pritiraj Mohanty$^{2}$, Richard A. Webb$^{3}$}
\author{Pascal Degiovanni$^{2,4}$, and Christopher B\"auerle$^{1}$}
\address{$^{1}$Institut  N\'{e}el, CNRS and Universit\'{e}
Joseph Fourier, B.P. 166, 38042 Grenoble, France}
\address{$^{2}$Department of Physics, Boston University, 590
Commonwealth Avenue, Boston, MA 02215, USA}
\address{$^{3}$USC Nano Center, University of South Carolina, Columbia SC 29208, USA}
\address{$^{4}$Laboratoire de Physique, \'{E}cole Normale
Sup\'{e}rieure de Lyon, 46 all\'{e}e d'Italie, 69364 Lyon Cedex 07,
France}

\date{\today}

\begin{abstract}
We review recent experimental progress on the saturation problem in
metallic quantum wires. In particular, we address the influence of
magnetic impurities on the electron phase coherence time. We also
present new measurements of the phase coherence time in ultra-clean
gold and silver wires and analyse the saturation of \tauphi in these
samples, cognizant of the role of magnetic scattering. For the
cleanest samples, Kondo temperatures below 1 mK and extremely-small
magnetic-impurity concentration levels of less than 0.08 ppm have to
be assumed to attribute the observed saturation to the presence of
magnetic impurities.
\end{abstract}
\maketitle

\section{Introduction}

The understanding of the ground state of an electron gas at zero
temperature is one of the major challenges in Condensed Matter Physics.
The conventional belief is that such a ground state is well described by
Landau's theory of Fermi liquids \cite{pines+nozieres}. This
description of physical and electrical properties of a metal starts from
the definition of quasiparticles, which are different from non-interacting
electrons. These quasiparticles decay with a finite lifetime at a finite temperature.
By definition, this lifetime has to diverge as temperature is lowered,
being infinite at zero temperature.

Within the framework of Landau's theory of Fermi liquids, all
scattering rates of an electron must vanish at zero temperature. As
shown by Altshuler and coworkers \cite{AAK_85}, Landau's theory also
holds in low dimensions and in presence of disorder.

The phase coherence time is one of the fundamental concepts in
mesoscopic physics. It is defined by the time an electron can travel
in a solid before losing its phase coherence and thus its quantum,
wave-like behaviour.  Such decoherence is usually due to inelastic
processes, like electron-phonon, electron-electron or
electron-photon collisions.  It has been shown by Altshuler and
coworkers \cite{AAK_85,AAK_82} that the phase coherence time of a
weakly disordered quantum conductor diverges at zero temperature as
electron-phonon, electron-electron and electron-photon interactions
all go to zero at zero temperature. Therefore, the Fermi liquid
description is also preserved for a disordered metal in low
dimensions, precisely because of the diverging phase coherence times
as temperature approaches absolute zero. Other dephasing mechanisms
such as electron scattering from magnetic impurity spins or
microwave radiation may generate a non-diverging or saturating
temperature dependence of the phase coherence time.

This intuitive picture, however, has been challenged by experiments
on metallic quantum wires which suggest that the phase coherence
time \tauphi saturates at very low temperature
\cite{mohanty_prl_97}. Following this work, it has been argued that
the observed saturation is indeed universal and intrinsic, and due
to electron-electron interactions in the ground state of the Fermi
liquid \cite{GZ_prl_98,GZS_jltp_02}. More recent studies of the
phase coherence time in metallic silver quantum wires
\cite{pierre_prb_03}, on the other hand, show a relatively good
agreement with the standard theory \cite{AAK_85,AAK_82}. In these
measurements, only small deviations of \tauphi compared with the
standard theory have been observed at the lowest temperatures, which
have been argued to be due to the presence of such a very small
amount of magnetic impurities.

In order to attribute the saturation of \tauphi in very clean
metallic quantum wires to the presence of a small amount of magnetic
impurities, it is important to have an adequate theory which
describes the underlying physics and enables a quantitative
comparison with the experimental data. Quite recently, there has
been tremendous progress in the understanding of the influence of
magnetic impurities on the phase coherence time, both experimentally
\cite{schopfer_prl_03,bauerle_prl_05,mallet_prl_06,birge_prl_06} and
theoretically \cite{zarand_prl_04,rosch_prl_06}. This new
understanding raises the question: Is it possible to reanalyse the
temperature dependence of \tauphi in very clean metallic wires to
determine whether the experimentally observed deviations from the
standard theory can be explained within this picture of magnetic
impurities.

The main purpose of this article is to review various experiments
which have addressed the influence of magnetic scattering on
electron coherence in metallic quantum wires. In the first part, we
briefly review the pioneering experiments which studied
electron coherence determined by magnetic scattering
\cite{bergmann_prl_87,haesendonck_prl_87,benoit_88,benoit_92}.

We then review recent measurements of electron coherence under high
magnetic fields $(\mu B >> k_B T)$ where the magnetic impurities are
polarised and therefore should not contribute to dephasing
\cite{benoit_88,benoit_92,birge_prl_02,mohanty_prl_03}.

In the last section, we present new data for the phase coherence time
in ultra clean gold and silver quantum wires, and compare them to
presently existing data on equally clean samples from other groups.
We then analyse the saturation observed in these samples assuming
that the apparent saturation of \tauphi is due to magnetic
impurities.

Our conclusion of this analysis is that, based on all presently
available measurements of the phase coherence time in very clean
metallic wires, it is hard to conceive that the apparent saturation
of \tauphi is solely due to the presence of an extremely small
amount of magnetic impurities.

\section{Review of Earlier Experiments on Magnetic Impurities}

As mentioned above, recent experiments invoke the presence of a
small amount of magnetic impurities as a possible source of the
frequently observed low temperature saturation of the phase
coherence time \cite{pierre_prb_03}. It is well known that the
coupling of magnetic impurities to the conduction electrons gives
rise to the well known Kondo effect \cite{kondo_64,hewson}. At
temperatures above the Kondo temperature $T_K$, the magnetic
scattering due to Kondo impurities leads to a very slow and an
almost temperature-independent contribution to the dephasing time
\cite{wei-bergmann_prb 89}. The magnetic contribution is maximal
around the Kondo temperature
\cite{bergmann_prl_87,haesendonck_prl_87} and decreases rapidly at
lower temperatures \cite{schopfer_prl_03,bauerle_prl_05}.
Consequently, if a metallic sample contains a small amount of
magnetic impurities with a very low Kondo temperature, the observed
temperature dependence of \tauphi would show saturation at
temperatures above $T_K$.

Already in the early days of weak localisation, many
experimentalists have observed a systematic saturation of the
electron phase coherence at low temperatures, when extracted from
low field magnetoresistance \cite{gershenzon,rosenbaum}. This
saturation has often been attributed to the presence of some
residual magnetic impurities \cite{bergmann_84}, however, without
any experimental verification.

To the best of our knowledge, the first experiment which clearly
demonstrated the strong influence of magnetic impurities on electron
coherence on the level of a few parts-per-million (ppm) in very
clean metallic samples has been carried out by Pannetier and
coworkers in 1985 \cite{pannetier_scripta}. The need for very long
phase coherence times in order to measure AAS oscillations in
two-dimensional networks \cite{pannetier_prl_84,pannetier_prb_85}
pushed the authors to seek extremely clean metals to obtain very
large values for the phase coherence length. The solution to the
problem was to thermally anneal the samples. The annealing process
oxidizes magnetic impurities, suppressing decoherence due to the
Kondo effect, and therefore leads to an increase of the phase
coherence length $L_\phi$. The phase coherence length of two gold
samples (before and after annealing) is shown in figure 1. One
clearly sees the enhancement of the phase coherence length due to
the annealing process \cite{comment_pannetier}.

\begin{figure}[h]
\centering{\includegraphics[width=7cm]{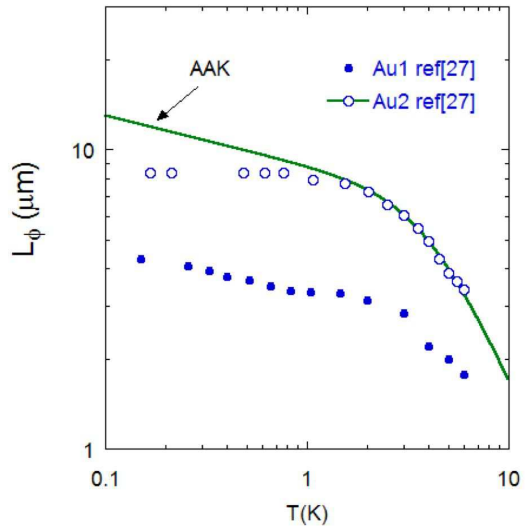}
\caption{\small Phase coherence length as a function of temperature
for an ultra-pure gold sample before ($\bullet$) and after ($\circ$)
annealing. The solid line corresponds to the theoretical expectation
within the AAK picture \cite{comment_network}. Data are taken from
ref. \cite{pannetier_scripta}.}}
  \label{pannetier}
\end{figure}

These experiments therefore clearly show that the presence of an
extremely small amount of magnetic impurities can lead to
substantial electron decoherence at low temperatures, as expected.
More importantly, even after the increase of the phase coherence
length due to the annealing process, the phase coherence length
still shows a relatively weak temperature dependence, much weaker
than the theoretically expected temperature dependence due to
electron-electron interaction \cite{AAK_82}.

A systematic study of the phase coherence time in the presence of
magnetic impurities has been pioneered by Bergmann et al.
\cite{bergmann_prl_87} and van Haesendonck et al.
\cite{haesendonck_prl_87}. Both groups have been able to measure the
Kondo maximum of the dephasing rate due to the scattering of
magnetic impurities in thin metallic films. Naturally, the
measurements have then been extended to temperatures below $T_K$ to
investigate the ground state of the Kondo problem. In this limit
($T\,\ll\,T_{K}$), Fermi liquid theory predicts a $T^2$ dependence
of the inelastic scattering rate \cite{nozieres_jltp_74}. These
experiments, however failed to observe this regime
\cite{bergmann_quest_FL}. Instead, a very slow temperature
dependence of the magnetic contribution to the dephasing rate has
been found below $T_K$ \cite{bergmann_T12}. This puzzle has remained
unresolved for almost 20 years. In connection to the saturation
problem \cite{mohanty_prl_97}, the questions of the Kondo ground
state has recently gained new interest
\cite{schopfer_prl_03,zarand_prl_04,bauerle_prl_05,rosch_prl_06,mallet_prl_06,birge_prl_06}.

\begin{figure}[h]
\centering{\includegraphics[width=7cm]{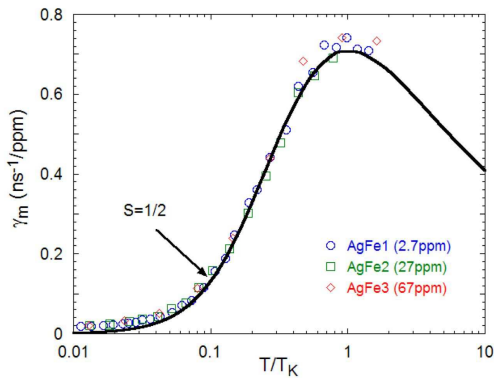} \caption{\small
Temperature dependence of the magnetic dephasing rate per impurity
concentration $\gamma_m$ of silver quantum wires doped with magnetic
iron impurities of different ion concentration. Data are taken from
ref. \cite{mallet_prl_06}. The solid line corresponds to the NRG
calculation of \tauphi for the single channel, S=1/2 Kondo model
\cite{rosch_prl_06}.}}
  \label{mallet}
\end{figure}

In a recent experimental study \cite{schopfer_prl_03}, a very
curious linear temperature dependence of \tauphi below the Kondo
temperature has been observed in AuFe quantum wires. An important
contribution to the understanding of this behavior came from theory,
where an exact calculation of the inelastic scattering time in Kondo
metals using Wilson's Numerical Renormalisation Group (NRG) approach
\cite{zarand_prl_04} was performed. The most interesting finding in
these calculations was the fact that in a temperature range
$0.1\,T_K<T<T_K$, the scattering rate due to magnetic impurities is
linear in T, perfectly consistent with the experimental findings
\cite{schopfer_prl_03,bauerle_prl_05}. Only for temperatures below
$0.1\,T_K$, the $T^2$ temperature dependence is expected. This
temperature regime has been explored experimentally only very
recently \cite{mallet_prl_06,birge_prl_06}. The experiment shows
\cite{mallet_prl_06} that the dephasing rate due to the magnetic
impurities is remarkably well described by the spin-1/2, single
channel model \cite{rosch_prl_06} as shown in figure \ref{mallet}.

\section{Experiments at high magnetic fields}

\begin{figure}[h]
\centering{\includegraphics[width=7cm]{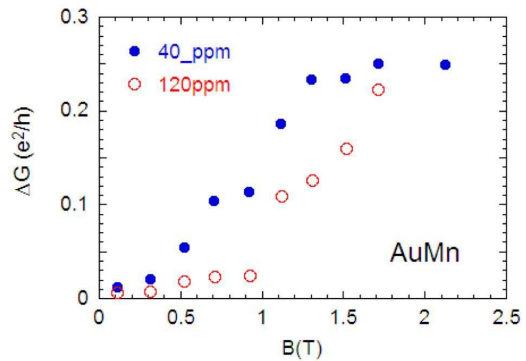}
  \caption{\small Amplitude of the Aharonov-Bohm conductance oscillations versus magnetic field
  for two gold samples with different Mn impurity concentration: 40ppm $(\circ)$ and 120 ppm $(\bullet)$.
  Data are taken from ref. \cite{benoit_92}.}}
  \label{benoit_AB}
\end{figure}

Another approach to investigate the influence of magnetic impurities on
the electron coherence involves phase coherent measurements
under strong magnetic fields. In this case, one can suppress the
effect of magnetic impurities by applying a sufficiently high
magnetic field in order to fully polarise the magnetic impurity
spins, and therefore study temperature dependence of phase coherence time
with no contribution from magnetic impurity spins. Because of the required
high fields in the order of teslas, low-field weak localisation measurements cannot be
used to extract the phase coherence time. However, measurements
of Aharonov Bohm (AB) oscillations and universal conduction
fluctuations (UCF) are possible.  Pioneering work on both, UCF and
AB oscillations in quasi-1D quantum conductors containing a small
amount of magnetic impurities (down to 40ppm) has been performed by
Beno\^\i t and coworkers in 1988 \cite{benoit_88}. These data sets are
shown in figure~\ref{benoit_AB}.

In this work, the authors clearly demonstrate that UCF as well as AB
oscillations increase considerably at fields larger than 1 Tesla,
showing the suppression of the Kondo effect due to the polarization
of the magnetic impurity spins. In the context of the present debate
on the low temperature saturation of $\tau_\phi$, these measurements
have been repeated recently on metallic samples containing more
dilute magnetic impurities~\cite{birge_prl_02}.
\bigskip

In this experiment, the authors measure the magnetoconductance of
copper rings and meander lines evaporated together. The phase
coherence time extracted from weak localization measurements exhibit
an anomalously strong saturation at low temperature for all samples
investigated. The authors claim, however, that no evidence of the
presence of Kondo impurities is detectable in the temperature
dependence of the resistivity.

For the two samples showing the most pronounced saturation of the
phase coherence time at low temperature, Aharonov-Bohm
magnetoconductance oscillations have been measured on a
ring structure at two different temperatures, $T\approx 40\,mK$ and
$100\,mK$. The magnetic field was swept up to $\approx 1.6\,T$,
\textit{i.e.} a field sufficient enough such that $\mu_{B}B\gg
k_{B}T$, with $\mu_{B}$ being the Bohr magneton. Superimposed on the
Universal Conductance Fluctuations, Aharonov-Bohm magnetoconductance
oscillations are clearly visible (see figure 3 of ref.
\cite{birge_prl_02}). The amplitude of these oscillations clearly
increases with magnetic field, as was observed by Beno{\^\i}t
\textit{et al.}~\cite{benoit_92} in their work on AuMn samples.

It is noteworthy that it is not straightforward to extract the phase
coherence length directly from the amplitude of the AB oscillations,
as the theoretical formula contains the phase coherence length both
in a prefactor (as a ratio of $L_{\phi}/L_{T}$, with $L_{T}$ the
thermal length) and in an exponential as the ratio of $L_{\phi}$
over the perimeter of the ring. Probably for this reason, the
authors do not try to extract $L_{\phi}$ from their data but rather
plot the amplitude of the AB oscillations as a function of the
magnetic field and fit their curves using the formula:
\begin{equation}
\Delta G_{AB}=C\frac{e^2}{h}\frac{L_{T}}{\pi
r}\sqrt{\frac{L_{\phi}}{\pi r}}\exp (-\frac{\pi r}{L_{\phi}})
\label{eq_deltaG}
\end{equation}
with $C$ a dimensionless constant, $e$ the charge of the electron,
$h$ the Planck constant, $r$ the radius of the ring and $L_{T}$ the
thermal length, $L_{T}=\sqrt{\hbar D\over k_{B}T}$ with $D$ the
diffusion constant and $T$ the temperature. $L_{\phi}$ is obtained
\textit{via} the formula $1/\tau_{\phi} = 1/\tau_{ee} + 1/\tau_{m}$,
$\tau_{ee}$ being the coherence time limited by the
electron-electron interactions and $\tau_{m}$ the inelastic
scattering time due to the presence of magnetic impurities. In the
limit $(B,T \gg T_K)$, the dephasing due to magnetic impurities is
given by:
\begin{equation}
\frac{\tau_{m}(B = 0)}{\tau_{m}(B)} = \frac{g\mu_{B}B/k_{B}T}{\sinh
{(g\mu_{B}B/k_{B}T)}} \label{tausf}
\end{equation}

There are obviously two unknown parameters in this problem, namely
${\tau_{m}(B = 0)}$ and $\tau_{ee}$. The authors assume that the
saturation of the coherence time at low temperature is due to
magnetic impurities; thus, they take ${\tau_{m}(B =
0)=\tau_{\phi}(B=0, T\rightarrow0)}$. As there is no direct
experimental way to determine $\tau_{ee}$, the authors take the
theoretical value given by the AAK formula (eq. \ref{AAK}; see
figure~4). Using these values in equations \ref{tausf} and
\ref{eq_deltaG}, they obtain a good agreement between their
experimental measurements of $\Delta G_{AB}(B)$ and the theoretical
expression (\ref{eq_deltaG}).

\begin{figure}[h]
\centering{\includegraphics[width=7cm]{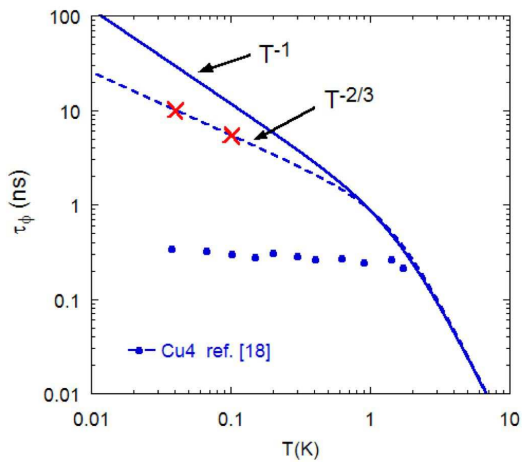}
  \caption{\small Phase coherence time as a function of temperature for a
  copper wire extracted from weak localisation measurements at low field and
  from AB oscillations at high fields. Data are taken from ref. \cite{birge_prl_02}.
  The dotted line corresponds
  to the AAK prediction for a quasi one dimensional wire and the solid lines to
  the AAK prediction for a AB ring geometry in the regime $L_{\phi} < L$, where $L$ is
  the perimeter of the ring \cite{LM_04,TM_05}. The two data points
  ($\times$) correspond to the theoretical values of the AAK theory for a quasi 1D wire,
  which have been used in ref. \cite{birge_prl_02} to fit the field dependence of the AB
  oscillations.}}
\label{birge}
\end{figure}

Note that, for the fitting procedure, the authors assume
that the temperature dependence of the phase coherence time is given
by the AAK formula for a quasi one dimensional (1D) wire:
\begin{equation}
\frac{1}{\tau_\phi}=\frac{1}{\tau_{ee}} + \frac{1}{\tau_{ep}} =
AT^{2/3} + BT^{3} \label{AAK}
\end{equation}
where the second term corresponds to the electron phonon
interaction. The $AT^{2/3}$ law holds only for the case of quasi 1D
wires. For the case of an Aharonov-Bohm ring, the correct
temperature dependence for $1/\tau_{ee}$ is in fact $AT^{1}$ (in the
limit $L > L_\phi$) as pointed out recently \cite{LM_04,TM_05}. If
one plots the correct temperature dependence as shown by the solid
line in figure~4, one clearly sees that the values for \tauphi used
to analyse the high field data ($\times$ in Fig.~4) lie much lower
than the theoretical expectation. This experiment hence shows that
the unusually strong dephasing observed in copper samples is
definitely due to the presence of magnetic impurities
\cite{comment_Cu}. However, it is clear that even under strong
magnetic fields, the dephasing time does \emph{not} recover the
theoretical expectation within the Fermi liquid picture for the AB
geometry \cite{LM_04,TM_05}. In a strict sense, this experiment
could even be interpreted as a \emph{support} for \emph{additional}
dephasing at zero temperature apart from magnetic impurities: if
under strong magnetic field, the \textquotedblleft
standard\textquotedblright description of dephasing due to
electron-electron interactions developed by AAK is not recovered,
there must be \emph{another} mechanism which leads to decoherence at
low temperatures \cite{birge_loc2002}.
\bigskip

Another experiment has been recently performed to clarify the role
of magnetic impurities in the low temperature saturation of the
phase coherence time \cite{mohanty_prl_03}. The fundamental
difference with the previous experiment is that here the authors
measure \emph{Universal Conductance Fluctuations} (UCF) in order to
extract $\tau_{\phi}$ and that they start from a \textquotedblleft
pure\textquotedblright~sample: although the phase coherence time
exhibits a clear saturation at low temperature, the resistivity as a
function of the temperature follows nicely the $1/\sqrt{T}$
dependence predicted for the electron-electron interaction
correction to the resistivity. For the UCF measurements, the authors
use short and narrow wires in order to maximize the signal. Two
different techniques are used to extract the phase coherence length:
first, the phase coherence length can be directly related to the
amplitude of the UCF $\delta G_{rms}$ \emph{via} the relation:
\begin{equation}
L_{\phi}=\frac{3\pi}{8}\frac{L^{3}}{{L_{T}}^{2}}\left(\frac{\delta
G_{rms}}{e^{2}/h}\right)^{2}
\end{equation}
where $L$ is the length of the sample and $L_{T}$ the thermal
length. An alternative way involves the measurement of the
correlation field $B_{c}$ defined as the field over which the
autocorrelation function $\left<G(B)G(B+\Delta B) \right>$ reaches
half of its zero field value. $L_{\phi}$ is then related to $B_{c}$
\emph{via}:
\begin{equation}
L_{\phi}=C\frac{h/e}{wB_{c}}
\end{equation}
where $w$ is the width of the sample and $C$ a numerical constant of
order $1$. In the experiment, the magnetic field is swept up to
$15\,T$, \emph{i.e.} $\mu_{B}B\gg k_{B}T$ as the temperature range
was from $39\,mK$ to $1\,K$. The phase coherence length extracted
from the UCF using both technics (amplitude and correlation field)
\emph{do exhibit a clear saturation of the phase coherence length}
at low temperature. From this experiment, the authors conclude that
the saturation of the phase coherence time in their samples is not
due to magnetic impurities but is rather intrinsic.

\begin{figure}[h]
\centering{\includegraphics[width=8cm]{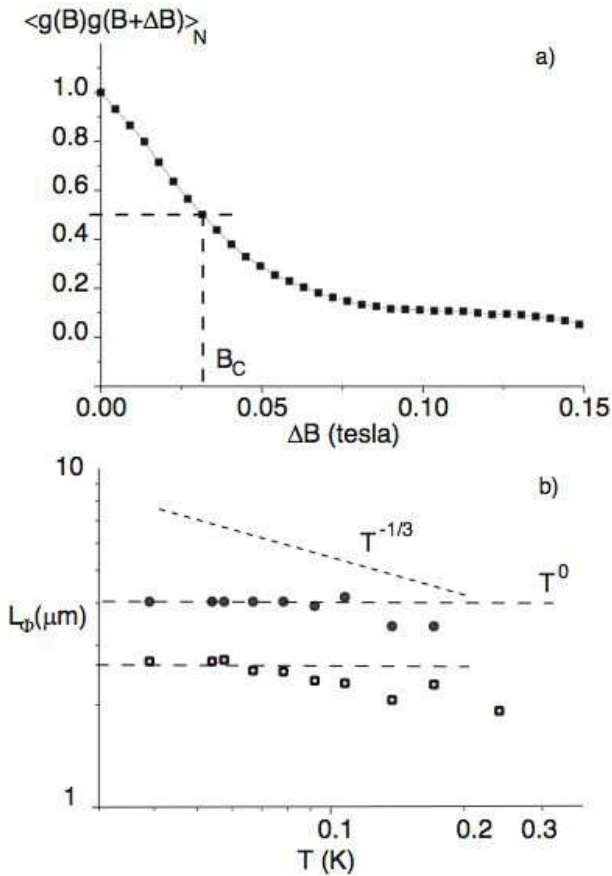}
  \caption{(a) Normalized autocorrelation function calculated from conductance
fluctuations above a field magnitude of 1 tesla. (b) Temperature
dependence of $L_\phi$ for two samples, determined from the
correlation field $B_c$, shows saturation. The two
quasi-one-dimensional gold wires have the following dimensions: 18
nm thick, 30 nm wide and 20 $\mu$m long, fabricated from gold with a
purity of 99.9995 $\%$. The data shown in Fig.5(a) is for a nanowire
with a resistance of 2390 $\Omega$. In Fig.5(b), $L_\phi$ for this
sample is represented by filled circles, while squares represent
data for a second sample with a resistance of 2886 $\Omega$.
Diffusion constant for both samples is approximately 0.005 m$^2$/s.
See Ref.\cite{mohanty_prl_03} for additional details.}}
  \label{mohanty}
\end{figure}

What have we learned from this set of experiments (Beno{\^\i}t
\emph{et al.}, Pierre \emph{et al.} and Mohanty \emph{et al.})? From
the two first, it is clear that magnetic impurities play an
important role in dephasing, and that a strong magnetic field can
suppress (or at least strongly reduce) such magnetic scattering.
However, none of these experiments have been able to measure the
temperature dependence of the phase coherence time under strong
magnetic field. Moreover, following recent theoretical progress,
one can state that in the AB oscillation \cite{birge_prl_02} as well
as in the UCF experiment \cite{mohanty_prl_03} the phase coherence
time determined at high fields is smaller than the one expected from
the standard Fermi liquid picture. In addition, the experiment of
Mohanty and Webb on pure wires seems to show that in their case the
phase coherence time does saturate at low temperature \emph{even
under strong magnetic field}: one can thus conclude that, whether or not
intrinsic, this saturation is not due to magnetic impurities. From
this panorama, it seems that magnetic impurities \emph{do} lead to
strong dephasing and \emph{partially} explain the low temperature
saturation of the phase coherence time, but these experiments do not
provide evidences strong enough to rule out an intrinsic saturation
as suggested by Mohanty \& Webb in their pioneering work. Let us
also mention that dephasing in much more disordered samples show
also an anomalously strong dephasing \cite{Lin_review} which is
presently not understood.

\section{Saturation of \tauphi in extremely pure metallic quantum wires}

Let us now discuss the temperature dependence of \tauphi in
extremely clean metallic quantum wires. In the absence of magnetic
impurities, the main decoherence mechanisms in mesoscopic wires is
assumed to be due to the electron-electron (ee) and electron-phonon
(ep) interaction. The temperature dependence of 1/\tauphi for a
quasi one-dimensional diffusive wire is then given by the following
expression \cite{AAK_82}:
\begin{equation}
\frac{1}{\tau_\phi}=\frac{1}{\tau_{ee}} + \frac{1}{\tau_{ep}} =
aT^{2/3} + bT^{3} \label{eq_AAK}
\end{equation}
\begin{displaymath}
\begin{array}{lll}
\frac{1}{\tau_{ee}} & = & a_{theo}\,T^{2/3}\\
 &  = & \bigg[\frac{\pi}{\sqrt{2}}\frac{R}{R_K}
\frac{k_B}{\hbar} \frac{\sqrt{D}}{L}\bigg]^{2/3} \, T^{2/3} \\
\\
\frac{1}{\tau_{ep}}\, & = & \,b\,T^{3}
\end{array}
\end{displaymath}

The ep-term dominates at temperatures above 1K whereas the
ee-interaction term is the leading contribution at lower
temperatures. Approaching zero temperature, the phase coherence time
should hence diverge with a power law proportional to $T^{-2/3}$. On
the other hand, almost all measurements on such wires show
deviations from this commonly believed picture on electron
dephasing. A few examples from different experimental groups are
shown in figure~6 and their electrical properties are summarized in
table \ref{table1}.

\begin{table}[h]
{\footnotesize
\begin{tabular}{@{}ccccccccc@{}}
\hline
{} &{} &{} &{} &{} &{} &{} &{} &{}\\[-1.5ex]
$Sample$ & $w $ & $t $ & $L $ & $R $ &$D $& $L_\phi^{max}$ & $\tau_\phi^{max}$ & Ref. \\[1ex]
$ $ & $(nm)$ & $(nm)$ & $(\mu m)$ & $(\Omega) $ &$(cm^{2}/s) $ & $(\mu m)$ & $ (ns) $ &  \\[1ex]
\hline
{} &{} &{} &{} &{} &{} &{} &{} \\[-1.5ex]
$Au^{2D}$ & $22\mu m$& $22$ & $44.3mm$ & $3.980$ & $135$ & $>28.7$ & $>61$ & $[39]$ \\[1ex]
$Au1$ & $120$ & $50$ & $450$ & $1218$ & $241$ & $>20.2$ & $>17$ & $[9]$  \\[1ex]
$Au_{MSU}$ & $90$ & $45$ & $175$ & $1080$ & $135$ & $>11.0$ & $>9$ & $[7]$ \\[1ex]
$Au_{EGB}$ & $90$ & $15$ & $100$ & $3.0$ & $135$ & $>3.7$ & $>1$ & $[40]$ \\[1ex]
$Ag1$ & $120$ & $50$ & $465$ & $2997$ & $105$ & $>15.8$ & $>24$ & this work \\[1ex]
$Ag6Na$ & $65$ & $45$ & $135$ & $1440$ & $115$ & $>10.2$ & $>9$ & $[7]$\\[1ex]
$Ag6Nc$ & $105$ & $55$ & $400$ & $1440$ & $185$ & $>20.2$ & $>22.0$ & $[7]$ \\[1ex]
\hline
\end{tabular}}
\vspace{0.5cm} \caption{Sample characteristics: $w, t, l, R,$
correspond to the width, thickness, length and the low temperature
electrical resistance, respectively. $D$ is the diffusion
coefficient, $L_\phi^{max}$ and $\tau_\phi^{max}$ the maximal values
of the phase coherence length (phase coherence time). Note that the
sample ($Au^{2D}$) \cite{webb_fortschritte_98} is two-dimensional
and sample $Au_{EGB}$ \cite{echternach_prb_93} consists of 21 narrow
Au strips, connected in parallel. The resistance for this sample
corresponds to the sheet resistance. \label{table1}}
\end{table}

In figure~6 we also plot the theoretical expectations, based on
equation~(6), as indicated by the solid lines. At sufficiently low
temperatures, all samples show deviations from the theoretically
expected prediction. Evaluating the theoretical parameter
$a_{theo}$, as listed in table \ref{table2}, we note that the
experimental value $a_{exp}$ determined from fitting the
experimental data to equation~(6) (solid lines in figure~6) is
\emph{always} lower than the expected theoretical value (dotted
line). Stated differently, this means that the measured dephasing time
is \emph{always} lower than the theoretically expected value.

\begin{figure}[h]
\centering{\includegraphics[width=7.5cm]{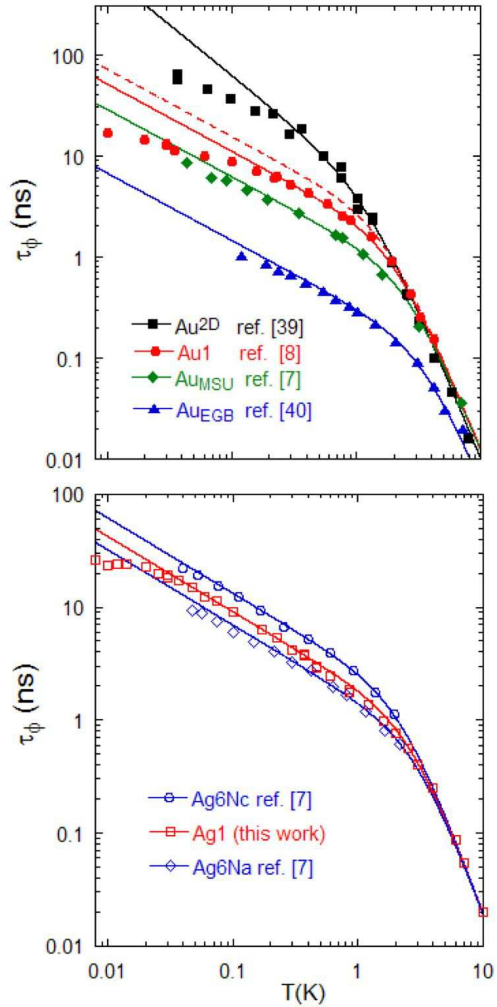}
  \caption{\small Phase coherence time as a function of temperature for
several metallic quantum wires. top: gold wires, bottom: silver
wires. The solid lines correspond to the theoretical expectations
within the AAK theory, using the experimentally determined
parameters $a_{exp}$ and $b_{exp}$ given in table 2. The dotted line
corresponds to the AAK theory for sample Au1 using the theoretically
expected prefactor from equation \ref{eq_AAK}.}} \label{clean_wires}
\end{figure}

Naturally, the observed deviations have been
associated with the presence of a tiny amount of magnetic impurities
which contribute to dephasing at the lowest temperatures. As we have
seen in section 2, magnetic impurities lead to strong dephasing
around the Kondo temperature due to inelastic magnetic scattering.
If magnetic impurities with a sufficiently low Kondo temperature are
for some unknown reason present in such a material, obviously one
would obtain an apparent saturation of \tauphi in the investigated
temperature range. To corroborate this picture, it is however
necessary to have a quantitative analysis of such
\emph{hypothetical} Kondo impurities and to see whether such an
assumption is reasonable.

Recent advances in the understanding of Kondo scattering in such
systems
\cite{bauerle_prl_05,mallet_prl_06,zarand_prl_04,rosch_prl_06} opens
the possibility to reanalyse existing data on \tauphi in extremely
clean metallic wires which show saturation and to verify whether
this saturation can be explained by the presence of a tiny amount of
magnetic impurities. As we have seen in section 2, the temperature
dependence of \tauphi in the presence of Kondo impurities is well
described by the $S=1/2$ single channel Kondo model, at least down
to temperatures of 0.1 $T_K$ \cite{mallet_prl_06}.

\begin{table}[h]
{\footnotesize
\begin{tabular}{@{}cccccccc@{}}
\hline
{} &{} &{} &{} &{} &{} &{} &{}\\[-1.5ex]
$Sample$ & $R  $ &$D $& $a_{theo} $ & $ a_{exp} $ & $ b_{exp} $\\[1ex]
$ $ & $ (\Omega) $ &$ (cm^{2}/s) $& $ (ns^{-1}K^{-2/3}) $ & $ (ns^{-1}K^{-2/3}) $ & $ (ns^{-1}K^{-3}) $\\[1ex]
\hline
{} &{} &{} &{} &{} &{} &{} \\[-1.5ex]
$Au^{2D}$ & $--$ & $135$ & $--$ & $0.16$ & $0.1$\\[1ex]
$Au1$  & $1218$ & $241$ & $0.28$ & $0.38$ & $0.08$\\[1ex]
$Au_{MSU}$ & $1080$ & $135$ & $0.4$ & $0.75$ & $0.08$\\[1ex]
$Au_{EGB}$ & $$ & $135$ & $--$ & $3.2$ & $0.16$\\[1ex]
$Ag1$ & $2997$ & $105$ & $0.37$ & $0.52$ & $0.05$\\[1ex]
$Ag6Na$ & $1440$ & $115$ & $0.55$ & $0.67$ & $0.05$\\[1ex]
$Ag6Nc$ & $1440$ & $185$ & $0.31$ & $0.35$ & $0.05$\\[1ex]
\hline
\end{tabular}}
\vspace{0.5cm} \caption{Fitting parameters extracted from equation
(6). Note that the experimental values of $ a_{exp} $ differ
slightly from the ones given in the original experiments. This comes
from the fact that we fit all data such that a good agreement is
obtained for the data at high temperatures $(T>200mK)$.
\label{table2}}
\end{table}

If magnetic impurities are somehow present in a metallic
quantum wire, this can be taken into account in the temperature
dependence of \tauphi by adding a third term in equation
\ref{eq_AAK}. Equation \ref{eq_AAK} reads then
\begin{equation}
\frac{1}{\tau_\phi}=\frac{1}{\tau_{ee}} + \frac{1}{\tau_{ep}} +
\frac{1}{\tau_{m}} \label{eq_NRG}
\end{equation}
\begin{displaymath}
\begin{array}{lll}
 & = & aT^{2/3} + bT^{3} + n_s * f(T/T_K)
\end{array}
\end{displaymath}
where $n_s$ is the impurity concentration in parts-per-million and
$f(T/T_K)$ the universal dephasing rate due to magnetic impurities,
recently calculated by NRG \cite{rosch_prl_06}. As $a$ and $b$ are
determined at relatively high temperatures ($T\,>\,200\,mK$; see
figure~6), the only adjustable parameters are the Kondo temperature
$T_K$ and the impurity concentration $n_s$. In the following we will
simulate the temperature dependence of the phase coherence time by
assuming the presence of a small amount of magnetic impurities of
concentration $n_s$ and a Kondo temperature $T_K$.

We will limit the analysis to our own data (sample Ag1 and Au1)
which cover the largest temperature range. The same analysis could
be done for all other existing data and one would arrive at the same
conclusions.

\begin{figure}[h]
\centering{\includegraphics[width=7cm]{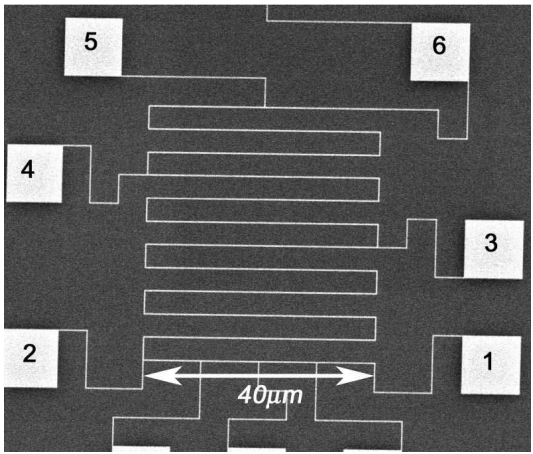}
  \caption{\small SEM picture of a typical quantum wire used in our experiments.
  The current is fed into contact 1 and the voltage is measured at contacts 2 and 5.
  Contact 6 is set to ground.}}
\label{sample}
\end{figure}

A typical SEM picture of one of our samples is shown in figure~7.
They are fabricated by standard electron beam lithography and
lift-off technique. The metal (Au or Ag) is evaporated directly onto
the Si wafer without adhesion layer. Special care has been taken for
the sample design such that there is no influence on the phase
coherence due to the two dimensional contact pads
\cite{prober_prl_88}. The resistance is measured via a standard
four-probe measurement technique using a lock-in amplifier in a
bridge configuration. This is necessary as the resistance variations
to be measured can be as small as $10^{-5}-10^{-6}$ of the absolute
value. The resistance measurement of the sample is made in a current
source mode: a voltage generated from a signal generator (typically
at a frequency of 13 Hz) is fed into the sample (for instance, at
contact 1) via a high resistance, typically of the order of
$10\,-\,100\,M\Omega$. The voltage across the quantum wire is then
measured between contact 2 and 5 (contact 6 is at ground) and
amplified by a home made preamplifier (EPC1) situated at room
temperature \cite{EPC1}. This voltage amplifier has an extremely low
noise voltage of about $0.4\,nV/\sqrt{Hz}$. We also take special
care of the stability of the amplification stage. Very stable
resistors $(1ppm / ^{\circ} C)$ are used for the current source. In
order to avoid radio-frequency (RF) heating due to external noise,
all measuring lines are extremely well filtered
\cite{comment_thermo}. At the sample stage we have an attenuation of
about $-400dB$ at 1K. We have tested the efficiency of the
attenuation by injecting a RF signal of frequency of 25\,GHz with an
amplitude of 25\,dB into the sample cell. Weak localisation
measurements at 30\,mK for sample Ag1, where there is still
agreement with the AAK theory, have not been affected by such a
procedure. All experiments have been performed in thermal
equilibrium which means that the applied voltage across the
\emph{entire} wire is kept such that the inequality $eV \leq k_B T$
is satisfied at all temperatures.

In order to verify that the electrons of the samples are cooled down
to the lowest measured temperatures, we measure the Altshuler-Aronov
correction to the resistivity at very low temperatures
\cite{AAK_85,mont_book}. As this correction is temperature
dependent, it allows us to determine the effective electron
temperature of our samples \emph{in situ}. For the measurements we
apply a magnetic field of 40\,mT in order to suppress weak
localization correction to the resistivity.

\begin{figure}[h]
\centering{\includegraphics[width=7.5cm]{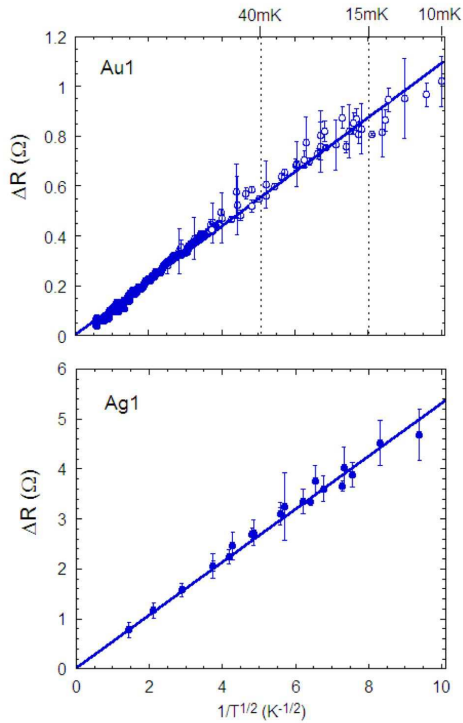}
  \caption{\small Resistance variation as a function of $1/\sqrt{T}$ for
  sample Au1 (top) and sample Ag1 (bottom). The solid line corresponds
  to a fit using equation \ref{lambda-sigma}.}}
\label{r_t}
\end{figure}

In figure~8 we plot the resistance correction as a function of
$1/\sqrt{T}$ for sample Au1. The resistance correction follows the
expected $1/\sqrt{T}$ temperature dependence down to 10\,mK. Fitting
the temperature dependence of the resistance correction to
\begin{equation}
\Delta R_{ee}(T) = \alpha_{exp}/\sqrt{T}
\end{equation}
(solid line in figure~8), we
determine $\alpha_{exp}$ and compare it to the predicted value
\cite{mont_book} of
\begin{equation}
\Delta R_{ee}(T) = 0.782\,\lambda_{\sigma}\,\frac{R^2}{R_K}\frac{L_T}{L}\\
=\lambda_{\sigma}\,\frac{\alpha_{theo}}{\sqrt{T}}
\label{lambda-sigma}
\end{equation}
where $L_T=\sqrt{\hbar D/k_B T}$ is the thermal length and
$R_K=h/e^2$. The parameter $\lambda_{\sigma}$ is a constant which
represents the strength of the screening of the interactions in the
metal under consideration. More precisely, it can be rewritten as a
function of a parameter $F$ which varies from $0$ for an unscreened
interaction to $1$ for a perfectly screened interaction. In one
dimension, one has $\lambda_{\sigma}=4-3F/2$. The values of
$\alpha_{exp}$ and $\alpha_{theo}$ are displayed in table
\ref{table3} \cite{comment_alpha}. From these values, we obtain
$\lambda_{\sigma}\approx 3.18$ for silver and $2.62$ for gold,
giving a value $F\approx 0.6$ for silver in good agreement with
theoretical values \cite{comment_copper}. For the gold sample,
however, we find $F\approx 0.9$, which is somewhat larger than
expected. Note, however, that in many experiments the coefficient
$F$ which can be deduced from the $R_{ee}(T)$
data\cite{pierre_prb_03} is obviously inconsistent with any
theoretical model as it is often negative! A careful analysis of the
available data is thus clearly needed in order to restore some
coherence between theory and experiment on this point of the e-e
interaction correction to the resistivity in metals.

\begin{table}[h]
{\footnotesize
\begin{tabular}{@{}ccccccccc@{}}
\hline
{} &{} &{} &{} &{} &{} &{} &{}&{}\\[-1.5ex]
$Sample$ & $R $ &$D $& $\alpha_{theo} $ & $\alpha_{exp}$&$\lambda_\sigma $ & $n_{s} $& $T_K $\\[1ex]
$ $ & $(\Omega) $ &$(cm^{2}/s) $& $(\Omega K^{-1/2}) $ & $(\Omega K^{-1/2})$ & $$ & $ (ppm) $ & $ (K) $ \\[1ex]
\hline
{} &{} &{} &{} &{} &{} &{} \\[-1.5ex]
$Au1$ & $1218$ & $241$ & $0.042$ & $0.11$ & $2.62$ & $<0.08$ & $\leq0.01$\\[1ex]
$Ag1$ & $2997$ & $105$ & $0.16$ & $0.35$ & $3.18$ & $<0.05$ & $\leq0.001$ \\[1ex]

\hline
\end{tabular}}
\vspace{0.5cm} \caption{Resistance correction to the resistivity due
to electron electron interactions . R is the electrical resistance
and $D$ is the diffusion coefficient. $n_s$ ($T_K$) is the maximal
ion concentration (maximal Kondo temperature) extracted from the
fits schown in figure 10 and 11.}\label{table3}
\end{table}

To determine the phase coherence time \tauphi we perform standard
magnetoresistance measurements as a function of temperature. Typical
measurements for sample Au1 and Ag1 are shown in figure~9.

\begin{figure}[h]
\centering{\includegraphics[width=7.5cm]{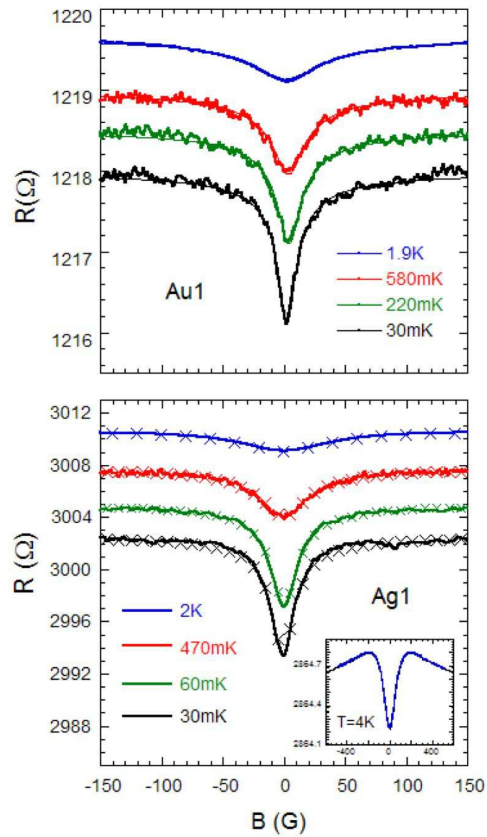}
 \caption{\small Magnetoresistance of gold sample Au1 (top) and silver sample Ag1(bottom) as
  a function of temperature. The solid lines are fits to equation \ref{eq_WL}.
  As the fits are almost indistinguishable from the data points,
  we have highlighted the fits for sample Ag1 with symbols $(\times)$.}}
\label{magneto}
\end{figure}

For a quasi one-dimensional wire the amplitude is essentially
proportional to \lphi/L, where L is the length of the wire and \lphi
is the phase coherence length. In addition, the width of the
magnetoresistance curve corresponds to the situation where exactly
one flux quantum $h/e$ is introduced into a loop of a surface of
$w\,\times\,$\lphi where $w$ is the width of the wire and \lphi the
phase coherence length. For the fitting of the magnetoresistance we
use the following expression \cite{hikami_80,imry_book,mont_book}:
\begin{displaymath}
\begin{array}{lll}
\frac{\Delta R}{R} & = &
\frac{2}{L}\frac{R}{h/e^{2}}\left\{\frac{3/2}
{\sqrt{\left[\frac{1}{l_{\phi}^2}+\frac{4}{3l_{so}^2}+\frac{1}{3}\left(\frac{wBe}{\hbar}\right)^2
\right]}}\right.\\
 &  & - \qquad \qquad
\left.\frac{1/2}{\sqrt{\left[\frac{1}{l_{\phi}^2}+\frac{1}{3}\left(\frac{wBe}{\hbar}\right)^2
\right]}}\right\}
\end{array}
\label{eq_WL}
\end{displaymath}
where B is the magnetic field and \lso the spin-orbit length. In the
low temperature limit (\lso\,$\ll$\,\lphi), the effect of the
spin-orbit scattering is to turn the weak localisation into weak
anti-localisation (minimum in the magnetoresistance)
\cite{bergmann_prl_82,bergmann_prb_83}. We also note that the above
formula is valid in the regime where \lphi\,$<\,L$.

From fitting the magneto-resistance to equation \ref{eq_WL} as
indicated by the solid lines in figure~9, we can extract the phase
coherence length $L_\phi$. For the fitting procedure we first
determine the spin orbit length $L_{so}$ at high temperatures,
typically several Kelvins where the spin orbit length \lso is of the
order of the phase coherence length (see inset of Fig. 9). For
silver (gold) samples we obtain typically $L_{so} \approx 500\,nm$
$(50\,nm)$. We then fix the spin orbit length and the only
adjustable parameter for the fitting procedure is $L_\phi$. The
phase coherence time \tauphi is then obtained from the relation
(\lphi$ = \sqrt{D \tau_{\phi}}$) and its temperature dependence for
sample Au1 and Ag1 is plotted in figure 10 (top) and 11 (top). We
also display the theoretical expectation based on the AAK theory as
indicated by the solid lines. As clearly seen, for both data sets
the experimentally measured value of \tauphi deviate substantially
from the AAK prediction at the lowest temperatures. To see whether
these deviations can be explained by the presence of a very small
amount of magnetic impurities, we simulate the temperature
dependence of \tauphi for the presence of a small amount of magnetic
impurities by making use of equation \ref{eq_NRG}. We first analyse
sample Au1 which shows stronger deviations from the AAK theory. For
the simulations we simply vary the \emph{possible} ion concentration
and the value of $T_K$. The blue (a), red (b) green (c) and black
(d) solid lines correspond to a simulation assuming
$T_K$\,=\,40\,mK, $T_K$\,=\,10\,mK, $T_K$\,=\,5\,mK, and
$T_K$\,=\,2\,mK, with an impurity concentration of
$n_{s}$\,=\,0.065\,ppm, $n_{s}$\,=\,0.08\,ppm, $n_{s}$\,=\,0.1\,ppm
and $n_{s}$\,=\,0.18\,ppm, respectively.

\begin{figure}[h]
\centering{
\includegraphics[width=8cm]{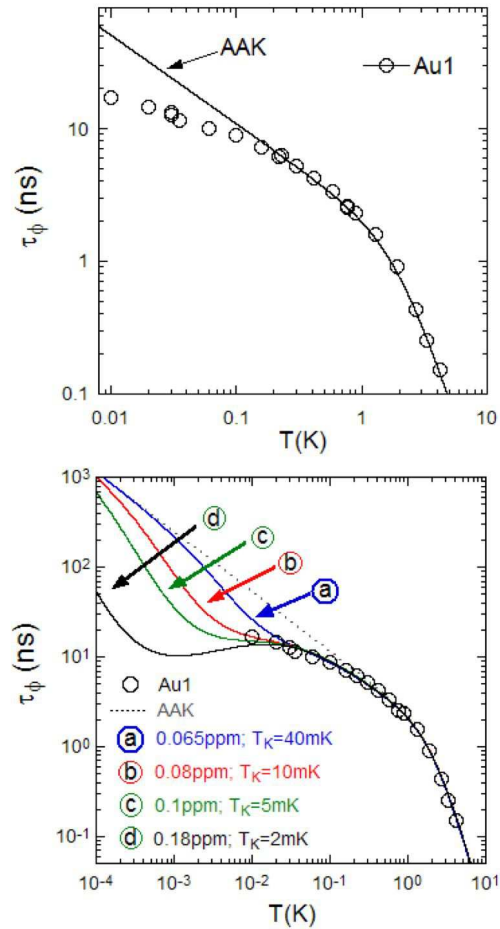}
  \caption{\small Phase coherence time as a function of temperature for
  sample Au1 ({\large $\circ$}). The solid green line corresponds to the AAK prediction,
 the black (a), red (b) and blue (c) solid lines correspond to the NRG
 calculation assuming $T_K=40\,$mK,
 $T_K=10\,$mK, $T_K=5\,$mK, and $T_K=2\,$mK respectively.}}
    \label{Au1_NRG}
\end{figure}

It is clear from our simulations that the temperature dependence of
\tauphi can only be  explained satisfactorily by the presence of
magnetic impurities with a Kondo temperature $T_K \leq$\,10\,mK and
with a concentration smaller than 0.08\,ppm. A possible magnetic
impurity with a Kondo temperature in this temperature range is Mn
($T_K\,\simeq$\,3mK) \cite{eska} and one could hence assign the
deviations at low temperatures to the presence of an ultra small
amount of Mn impurities. The question is whether this assumption is
reasonable and realistic to explain the observed deviations compared
to the standard AAK picture?

As the issue of low temperature saturation is rather polemic and
since the discussions are sometimes far from being scientific, we
would like to take an objective stand to the saturation problem. In
the following, we will give several comments to the above
interpretation and point out several inconsistencies with this
interpretation. The aim is not to give any definite answer to the
dephasing problem, but to present experimental facts in a very
objective way, and offer possible interpretations of the experimental
findings.

If one assumes the presence of magnetic impurities with a Kondo
temperature below the measurement temperature, this will lead to an
almost temperature independent scattering rate for $T\geq
T_K$\footnote{the exact temperature dependence is $\propto
ln^2(T/T_K)$ for $T \gg T_K$ and approaches a constant at $T_K$.}.
Any experimentally observed saturation of \tauphi can therefore
always be assigned to magnetic impurities with an unmeasurably low
Kondo temperature. One could also argue, it is curious that for the
case of gold wires, the observed temperature dependence of \tauphi
can only be described satisfactorily by assuming the presence of one
specific magnetic impurity with a Kondo temperature below the
measuring temperature $T\leq $\,10mK, whereas it is known that the
dominant magnetic impurity in gold is iron \cite{laborde_these}. If
we assume an additional iron concentration of the same order as, for
instance Mn, the temperature dependence of \tauphi does not
satisfactorily describe the experimental data \cite{bauerle_prl_05}.

\begin{figure}[h]
\centering{
\includegraphics[width=8cm]{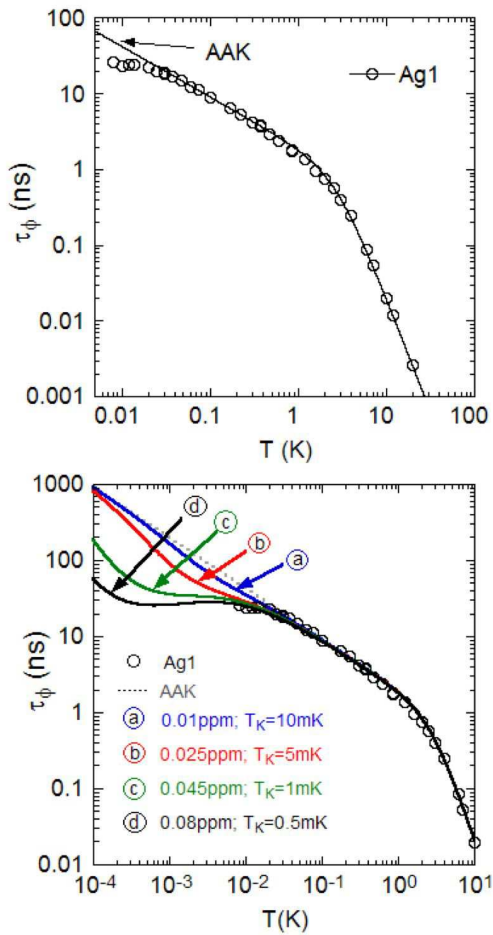}
  \caption{\small Phase coherence time as a function of temperature for
  sample Ag1 ({\large $\circ$}).
 The solid green line corresponds to the AAK prediction,
 the black (a), red (b) and blue (c) solid lines correspond to the
 NRG calculation assuming $T_K=5\,$mK,
 $T_K=1\,$mK, $T_K=0.5\,$mK, and $T_K=0.1\,$mK, respectively.}}
    \label{Ag1_NRG}
\end{figure}

The discrepancy actually gets worse if we analyse our results for
the ultra pure silver wire. This is displayed in figure~11. The
green solid line corresponds again to the AAK prediction. At
temperatures below 30\,mK, our data deviate again significantly from
the AAK prediction. The black (a), red (b) and blue (c) solid lines
correspond to a simulation assuming $T_K$\,=\,10\,mK,
$T_K$\,=\,5\,mK, $T_K$\,=\,1\,mK and $T_K$\,=\,0.5\,mK, with an
impurity concentration of $n_{s}$\,=\,0.01\,ppm,
$n_{s}$\,=\,0.025\,ppm, $n_{s}$\,=\,0.045\,ppm, and
$c_{imp}$\,=\,0.08\,ppm, respectively. It is clear from our
simulations that only magnetic impurities with a Kondo temperature
\emph{smaller} than 1\,mK and with a concentration smaller than
$0.08\,ppm$ describe satisfactorily the experimental data. This is
indeed surprising, as the lowest known Kondo temperature for a Ag
host is around 20-40 mK~\cite{daybell_review,rizutto}.

For the sake of completeness, we would like to point out that
disorder introduces a modification for the spin relaxation rate a
low temperatures. Ketteman and collaborators have suggested that
magnetic impurities with very low Kondo temperature may be present
in diffusive conductors since the Kondo temperature is strongly
dependent on the local DOS \cite{kettemann_jetp_06}. Recent
numerical simulations for 2D disordered metals suggest that the
dispersion of Kondo temperatures induced by non-magnetic disorder
can enhance the spin relaxation rate at low temperatures compared to
the case of a clean sample \cite{ketteman_2006}. On the other hand
the distribution of $T_K$ seems to be too small for the investigated
samples. This scenario, however, requires a more systematic analysis
of the distribution of Kondo temperatures in quasi 1D systems, in
particular their dependence on the aspect ratio.

\section{Comparison with theory}

From what we have seen so far, it seems clear that  \emph{magnetic
impurities do play an important role in dephasing}, and that this
dephasing can be understood correctly within the framework of the
Kondo Physics. However, it also seems clear that an \emph{intrinsic}
saturation of the phase coherence time at low temperature cannot be
ruled out from the sole experimental evidence available so far.
Although controversial, a theoretical model has been developed by
Golubev and Zaikin \cite{GZ_prl_98}. This model is an alternative to
the standard AAK theory which claims that intrinsic decoherence at
zero temperature is present even in a Fermi liquid type of
framework. In this section, we would like to present the two models
from a more formal point of view, stressing where the two approaches
converge and where they diverge. However, we will not attempt here
to compare these two models with the presented experiments. An
interesting comparison of the theory of Golubev and Zaikin with
experiments can be found in ref. \cite{GZ_physica_E_07}
\bigskip

In condensed matter physics, understanding the behavior of an
electron in its electromagnetic environment is an old problem. One
of the central ideas is to integrate over the environmental degrees of
freedom, thus encapsulating the effect of the environment on the
particle into an influence functional. This method, originally due
to Feynman \cite{Feynman_Vernon}, assumes that the environmental
degrees of freedom are distinct from the electronic ones. This is
obviously the case when they are of different nature (phonons,
magnetic impurities). This is also the case when the test particle
is indistinguishable from the environmental particles provided
tunneling between the particle and the environment can be neglected
as in the case of electrons of an external gate. A clear
separation between system and environmental degrees of freedom is
then restored and the test particle only feels electromagnetic
fluctuations created by charge and current density fluctuations in
the gate.

In a pioneering effort, Guinea and coworkers have studied the
influence of a normal Fermi liquid on
 charged particles \cite{guinea_prl_53}, the dephasing of electrons
in a diffusive wires under the influence of a nearby metallic gate
\cite{guinea_prb_71} and the lifetime of electrons in ballistic
quantum dots \cite{guinea_prb_70}.

\medskip

On the other hand, the theoretical debate triggered by experimental
results of Mohanty {\it et al} \cite{mohanty_prl_97} concerns the
coherence properties of an electron in a metal under the influence
of electromagnetic fluctuations created by the other electrons.
Because the test particle and the environmental electrons belong to
the same electronic fluid, the standard influence functional
approach is expected to break down since it does not incorporate
exchange effects (Pauli principle). One has to deal with a fully
interacting system of electrons coupled through the Coulomb
interaction and possibly interaction with real photons.

\medskip

Before turning to this difficult problem, let us recall that in 1D
metallic systems, Coulomb interactions eventually lead to the
breakdown of the Fermi liquid theory. Nevertheless, the effective
theory describing the low energy behavior of the electronic fluid is
well known: it is the Luttinger liquid theory \cite{haldane_1981}.
Its stability under the presence of electric potential fluctuations
induced by an external gate has been studied
\cite{cazalilla_prl_2006} and dissipation driven transitions towards
two different phases have been predicted for sufficiently strong
repulsive interactions. In the Luttinger liquid phase, decoherence
induced by phonons or by real photons can indeed be computed exactly
\cite {degio_prb_62}. Therefore, the Luttinger liquid theory
provides a very interesting theoretical framework for studying the
behavior of a dissipative strongly-interacting electronic quantum
system, showing how coupling to environmental degrees of freedom can
change the fluid's properties or even trigger phase transitions and
induce decay of Schr\H{o}ˆdinger cat states.

In the case of a 3D, 2D or quasi-1D diffusive conductor, the Fermi
liquid theory is the only known low-energy effective theory for the
electronic fluid. Consequently the above questions are much more
difficult to address than in the case of Luttinger liquid. This is
exactly why raising the possibility of an intrinsic saturation of
electron dephasing rate at very low temperature is so important: it
would be a sign of the breakdown of the Fermi liquid picture which describes
the ground state of metals, and associated low-energy properties.
This obviously explains the intensity and the relevance of the theoretical
debate raised by the work of Golubev and Zaikin
\cite{GZ_prl_98,GZ_prb_99}.

\medskip

In their work, Golubev and Zaikin (GZ) claim two results: (1) the
correct derivation of an influence functional formalism for the
dynamics of an electron in a diffusive conductor taking into account
electromagnetic fluctuations created by the other electrons and the
Pauli principle, and (2) the claim that these intrinsic
electromagnetic quantum fluctuations lead to a saturation of the
dephasing time of an electron in a diffusive conductor at zero
temperature. After years of polemics, their derivation of an
influence functional taking into account the Pauli principle has
been acknowledged first by Eriksen {\it et al} \cite{Eriksen:2001-1}
who used this tool to compute the renormalization of the effective
mass for composite fermions and then by J.~von~Delft
\cite{von_delft_long}. The GZ functional influence is an
\emph{important result} by itself since it potentially enables to
account precisely the effects of quantum electromagnetic
fluctuations on the electron dynamics whereas the original approach
by AAK \cite{AAK_82} only dealt with the effect of classical
fluctuations (thermal Nyquist noise).

\medskip

Because of these recent developments, the controversy has now been
shifted to the second claim by Golubev and Zaikin since
von~Delft and his collaborators claim that the AAK results
can be in fact recovered from the functional influence approach. They
also argue that a simple Fermi Golden rule computation shows how
the Pauli principle prevents intrinsic decoherence at vanishing
temperature (see section V of \cite{von_delft_short1}). Some time
ago, Golubev and Zaikin had already argued that their approach leads
to terms that could not be obtained through the Fermi Golden rule
approach and that those terms are precisely responsible for zero
temperature dephasing \cite{Golubev_PRB_62}. But this criticism
should not be confused with the point made by Montambaux and
Akkermans \cite{Montambaux_PRL_95} who showed that a proper
treatment of the infrared cutoff in the Fermi Golden rule
computation leads to a non exponential phase relaxation of an
electron in a quasi-one dimensional disordered conductor treated
within the AAK theory \cite{AAK_82}. Also, more precise computations
presented in \cite{von_delft_short1} which take into account this
effect also lead to the same non-exponential behaviour of decay
functions for quasi-one dimensional conductors. The real debate has
to do with the effect of high-frequency environmental modes (with
respect to $k_{B}T/\hbar$) within the influence functional approach.
The basic argument against Golubev and Zaikin's results (2)
is that, at vanishing temperature, an electron at energy
$\epsilon $ above the Fermi surface cannot relax more than
$\epsilon$ because of the Pauli principle. Thus, the Pauli principle
provides an ultra-violet cutoff in the frequency integrals that
determine decay functions and this forces the dephasing rate to
vanish at zero temperature. Nevertheless, the von~Delft {\it et al.}
analysis has then been severely criticized by Golubev and Zaikin
\cite{GZ_comment} who argue that von Delft and coworkers use a different
influence functional which violates fundamentally important principles such as
causality and detailed balance. At this point, it is not the purpose
of the present paper to close this theoretical debate. Instead, we
would like to clarify where the debate has been relocated and stress
the difficulty of this theoretical problem.

\medskip

First of all, from a theoretical perspective, the problem is far
from being trivial. Because we are dealing with a many body
interacting system, the GZ formalism is indeed a self consistent
approach to the electron dynamics: the influence functional itself
depends on the properties of the electronic fluid and is basically
derived in an operatorial form. It must be understood that
evaluating and using this influence functional within a path
integral formalism is nothing but straightforward. The influence
functional involves propagators with effective Hamiltonians
containing an effective one-particle density operator which is
expected to be non local (in position). To deal with this problem,
GZ rely on a phase space path integral. Eriksen {\it et al} have
basically followed the same track and used the same method
\cite{Eriksen:2001-1}. On the other hand, von Delft
\cite{von_delft_long} rely on a configuration space path integral
(position only). In their comment  \cite{GZ_comment}, GZ point out
that the resulting influence functional is different from theirs. In
the same comment, they also mention several serious difficulties
with the von Delft evaluation of the influence functional and thus
seriously question the validity of von Delft's subsequent results.
As mentioned before, resolving this controversy is beyond the scope
of the present paper.

On the other hand, it is important to understand where the theoretical
debate has shifted. Both Golubev and Zaikin, and von Delft and his
collaborators agree on the existence of an influence functional
which enables describing the many body problem within an effective
one particle formalism. They also agree on the definition of the
influence functional under operatorial form. But their work diverge
on the implementation of this object within a path integral
formalism and the evaluation of the path integrals, needed to compute
the weak localization correction to the conductivity.

Golubev and Zaikin rely on a phase space path integral which enables
them to perform an explicit integration over the auxiliary
fluctuating fields introduced by the Hubbard-Stratonovich
transformation. A word of caution, phase space
path integrals can be flawed when evaluating a propagator associated
with an Hamiltonian whose kinetic term is not quadratic. As stressed
before, the Golubev-Zaikin influence functional precisely involves
some effective Hamiltonians which are not quadratic in the
particle's momentum. The well known and already less pathological
example of a particle moving on a Riemannian manifold provides an
illustration of these problems: the quantum mechanical amplitude is
not given by the naive phase space path integral. Explicit
evaluation of the phase space path integral requires an extra factor
(reflecting the metric over path's space induced by the Riemannian
metric) which usually cancels divergences arising in the naive phase
space integral. Moreover, it requires going back to a
discretized form which ultimately amounts to a choice of
quantization \cite{Zinn-book}.

On the other hand, von Delft uses the usual configuration space path
integral which certainly provide a safer path with respect to the
above mentioned concerns. But the presence of a non local operator
in the effective Hamiltonians is a serious challenge to its explicit
evaluation. Another way to avoid this complication by getting around
the use of phase space path integrals for non quadratic Hamiltonians
has been proposed by Golubev and Zaikin in Ref. \cite{GZ_JLTP_132}
but it is not yet totally clear to us which of these lines of work
provides a correct quantitative evaluation of intrinsic electron
decoherence at very low temperature.
\medskip

To summarize the theoretical debate suggested above, it seems
reasonable to claim that the Golubev and Zaikin influence functional
{\em defined in an operatorial way} is correct and properly
incorporates the effect of the Pauli principle as it has been
demonstrated by von Delft. But, the precise use and evaluation of this
influence functional within a path integral approach is a rather
subtle question whose understanding will ultimately lead to a
solution to the controversy over the discrepancy between Golubev and
Zaikin and AAK results.

\section{Conclusion}

In this article, we have reviewed recent experiments on dephasing in
metallic quantum wires in regards to magnetic impurities. From the
data we have presented, it is clear that magnetic impurities lead to
strong dephasing at low temperatures. However, all experiments
presented here strongly indicate that there must be an additional
mechanism which leads to dephasing at low temperature \emph{apart
from magnetic impurities}. Let us therefore emphasize that at
present there is no experimental evidence \emph{against} zero
temperature saturation. In our opinion, it does not make much sense
to look for cleaner and cleaner materials to see whether there is a
saturation of \tauphi at the lowest temperatures: indeed, the
controversial theory by Golubev and Zaikin \cite{GZ_prl_98} predicts
a $D^3$ dependence of the zero temperature saturation time $\tau_0$
\cite{GZS_jltp_02}, with $D$ being the diffusion coefficient. Going
to cleaner and cleaner materials would obviously increase the
diffusion coefficient and the saturation, \textit{if existing},
would never be observable at presently accessible temperatures.
Instead one should try to look for a system where the diffusion
coefficient can be varied over possibly one decade in $T$ without
changing the purity of the system. This was attempted in the
original paper by Mohanty and Webb.

We acknowledge helpful discussions with P. Simon, G. Zar\'and, L.
Borda, A. Rosch, T. Costi, A. Zawadowski, S. Kettemann, C. Texier,
G. Montambaux, M. Lavagna and D. Feinberg. We are indebted to the
Quantronics group for the silver evaporation. L.S. acknowledges
support from the \textit{Institut Universitaire de France}. P.D.
ackowledges support from the Condensed Matter Theory visitors
program at Boston University and C.B. support from the CNRS-DREI
(No. 4024). This work has been supported by the European Comission
FP6 NMP-3 project 505457-1 Ultra-1D and ANR-PNANO \textit{QuSpin}.
P.M. acknowledges support from the National Science Foundation
(CCF-0432089).


\begin{thebibliography}{99}
\bibitem{comment_pannetier} For sample Au1 one can actually see a small inflection of \lphi
around a temperature of 1\,K. This inflection is probably due to the
presence of a small amount of Fe impurities which have a Kondo
temperature in this temperature range. After annealing this
inflection disappears and the phase coherence length increases.
However, the phase coherence length still does not reach the
theoretical prediction.
\bibitem{EPC1} the EPC1 voltage amplifier is developped by the electronic service of the
CRTBT.
\bibitem{comment_alpha} note that here we defined $\alpha_{theo} =
\lambda_{\sigma} * \alpha_{exp}$.
\bibitem{comment_copper} For copper, for which the Thomas-Fermi length is
similar to the one of gold, $F=0.6$ \cite{mont_book}.
\end{thebibliography}
\end{document}